\documentclass{jpsj-suppl}

\title{
	Equation of state for neutron stars: \\ 
	Hyperon mixing in SU(3) flavor symmetry
}

\author{
	Tsuyoshi \textsc{Miyatsu}$^{1}$,  
	Myung-Ki \textsc{Cheoun}$^{1}$  
	and  
	Koichi \textsc{Saito}$^{2}$
}

\inst{
	$^{1}$Department of Physics,  Soongsil University,  Seoul 156-743,  Korea  \\
	$^{2}$Department of Physics,  Faculty of Science and Technology,  \\
	Tokyo University of Science (TUS),  Noda 278-8510,  Japan,    
	and \\
	J-PARC Branch, KEK Theory Center,  \\
	Institute of Particle and Nuclear Studies,  KEK,  Tokai 319-1106,  Japan
}

\email{tmiyatsu@ssu.ac.kr}

\recdate{July 12, 2013}

\abst{
	Using various kinds of relativistic mean-field models as well as the quark-meson coupling model,  
	we study in detail the properties of neutron stars.  We find that the equation of state in SU(3) flavor symmetry  
	can support a neutron star with mass of $(1.8 \sim 2.1) M_{\odot}$  
	even if hyperons exist inside the core of a neutron star. 
}

\kword{
	neutron stars,  
	relativistic mean-field theory,  
	SU(3) flavor symmetry
}

\begin{document}
\maketitle

\section{Introduction}

Recently, neutron stars with the mass around $2M_{\odot}$ ($M_{\odot}$: the solar mass) have been reported.  
For example,  the binary millisecond pulsar,  J1614-2230,  has the mass of $1.97 \pm 0.04 M_{\odot}$~\cite{Demorest:2010bx}, and the mass of pulsar,  J0348+0432,  is estimated to be $2.01 \pm 0.04 M_{\odot}$~\cite{Antoniadis:2013}.  
However,  it is difficult to explain such heavy neutron stars by the equation of state (EoS) which have been calculated in mean-field theory so far,  because the degrees of freedom of hyperons ($Y$) make the EoS very soft and the maximum mass of a neutron star is thus reduced.  
To solve this ``hyperon puzzle'',  we examine in detail the extension of SU(6) spin-flavor symmetry based on the quark model to SU(3) flavor symmetry in determining the couplings of the isoscalar,  vector mesons to the octet baryons.  

\section{Relativistic Mean-Field Models and SU(3) Flavor Symmetry}

We adopt the extended version of the relativistic mean-field (RMF) model which includes not only the $\sigma$, $\omega$ and $\vec{\rho\,}$ mesons but also the strange mesons,  namely the isoscalar,  Lorentz scalar ($\sigma^{\ast}$) and vector ($\phi$) mesons.  
The $\sigma^{\ast}$ and $\phi$ mesons are supposed to be predominantly composed of $\bar{s}s$ quarks.  
The Lagrangian density is thus chosen to be
\begin{eqnarray}
	\mathcal{L}
	&=& \sum_{B}\bar{\psi}_{B}\left[
    i\gamma_{\mu}\partial^{\mu}
	- M_{B}^{\ast}\left(\sigma,\sigma^{\ast}\right)
	- g_{\omega B}\gamma_{\mu}\omega^{\mu}
	- g_{\phi B}\gamma_{\mu}\phi^{\mu}
	- g_{\rho B}\gamma_{\mu}\vec{\rho}^{\,\mu}\cdot\vec{I}_{B} \right]\psi_{B}
	\nonumber \\
	&+& \frac{1}{2} \left( \partial_{\mu}\sigma\partial^{\mu}\sigma-m_{\sigma}^{2}\sigma^{2} \right)
	+ \frac{1}{2} \left( \partial_{\mu}\sigma^{\ast}\partial^{\mu}\sigma^{\ast}
	- m_{\sigma^{\ast}}^{2}\sigma^{\ast2} \right)
	\nonumber \\
	&+& \frac{1}{2}m_{\omega}^{2}\omega_{\mu}\omega^{\mu} - \frac{1}{4}W_{\mu\nu}W^{\mu\nu}
	+ \frac{1}{2}m_{\phi}^{2}\phi_{\mu}\phi^{\mu} - \frac{1}{4}P_{\mu\nu}P^{\mu\nu}
	+ \frac{1}{2}m_{\rho}^{2}\vec{\rho}_{\mu}\cdot\vec{\rho}^{\,\mu}
	- \frac{1}{4}\vec{R}_{\mu\nu}\cdot\vec{R}^{\mu\nu}
	\nonumber \\
	&-& U_{NL}(\sigma,\omega^{\mu},\vec{\rho}^{\,\mu})
	+ \sum_{\ell}\bar{\psi}_{\ell}\left[i\gamma_{\mu}\partial^{\mu}-m_{\ell}\right]\psi_{\ell} \ ,
  \label{eq:total-Lagrangian-density}
\end{eqnarray}
with $W_{\mu\nu}=\partial_{\mu}\omega_{\nu}-\partial_{\nu}\omega_{\mu}$, $P_{\mu\nu}=\partial_{\mu}\phi_{\nu}-\partial_{\nu}\phi_{\mu}$,  $\vec{R}_{\mu\nu}=\partial_{\mu}\vec{\rho}_{\nu}-\partial_{\nu}\vec{\rho}_{\mu}$, and $\psi_{B (\ell)}$ being 
the baryon (lepton) field.  The lepton mass is denoted by $m_{\ell}$, and  $\vec{I}_B$ is the isospin matrix for baryon $B$.  
The sum $B$ runs over the octet baryons,  $N$ (proton and neutron),  $\Lambda$,  $\Sigma^{+,0,-}$ and $\Xi^{0,-}$,  and the sum $\ell$ is for the leptons,  $e^{-}$ and $\mu^{-}$.  
The $\omega$-,  $\phi$- and $\rho$-$B$ coupling constants are respectively denoted by $g_{\omega B}$,  $g_{\phi B}$ and $g_{\rho B}$.  
In addition,  the nonlinear (NL) potential is given by
\begin{equation}
	U_{NL}(\sigma,\omega^{\mu},\vec{\rho}^{\,\mu})
	= \frac{1}{3}g_{2}\sigma^{3}
	+ \frac{1}{4}g_{3}\sigma^{4}
	- \frac{1}{4}c_{3}\left(\omega_{\mu}\omega^{\mu}\right)^{2}
	- \Lambda_{\omega\rho}
	\left( \omega_{\mu}\omega^{\mu}\right)\left(\vec{\rho}_{\mu}\cdot\vec{\rho}^{\,\mu} \right) \ .
	\label{eq:Lagrangian-NL}
\end{equation}

In Quantum Hadrodynamics (QHD) where the baryons are treated as point-like objects,  the effective baryon mass, $M_{B}^{\ast}$,  in matter is expressed as
\begin{equation}
	M_{B}^{\ast}\left(\sigma,\sigma^{\ast}\right)
	= M_{B} - g_{\sigma B}\sigma - g_{\sigma^{\ast}B}\sigma^{\ast} \ ,
	\label{eq:effective-mass-QHD}
\end{equation}
where $M_{B}$ is the mass in vacuum,  and $g_{\sigma B}$ and $g_{\sigma^{\ast}B}$ are the $\sigma$- and $\sigma^{\ast}$-$B$ coupling constants,  respectively.  
In contrast,  in the quark-meson coupling (QMC) and chiral quark-meson coupling (CQMC) models,  the coupling constants,  $g_{\sigma B}$ and $g_{\sigma^{\ast}B}$,  depend on the $\sigma$ and $\sigma^{\ast}$ fields,  which reflects the variation of baryon structure in matter. 
Thus,  in-medium baryon mass can be written as~\cite{Miyatsu:2013yta}
\begin{equation}
	M_{B}^{\ast}\left(\sigma,\sigma^{\ast}\right)
	= M_{B} - g_{\sigma B}(\sigma)\sigma - g_{\sigma^{\ast}B}(\sigma^{\ast})\sigma^{\ast} \ ,
	\label{eq:effective-mass-QMC}
\end{equation}
with the field-dependent coupling constants
\begin{equation}
	g_{\sigma B}(\sigma)
	= g_{\sigma B}b_{B}\left[1-\frac{a_{B}}{2}\left(g_{\sigma N}\sigma\right)\right] \ ,
	\hspace{0.5cm}
	g_{\sigma^{\ast}B}(\sigma^{\ast})
	= g_{\sigma^{\ast}B}b_{B}^{\prime}
	\left[1-\frac{a_{B}^{\prime}}{2}\left(g_{\sigma^{\ast}\Lambda}\sigma^{\ast}\right)\right] \ ,
	\label{eq:cc-Lorentz-scalar}
\end{equation}
where $g_{\sigma N}$ and $g_{\sigma^{\ast}\Lambda}$ are respectively the $\sigma$-$N$ and $\sigma^{\ast}$-$\Lambda$ coupling constants at zero density, and 
we introduce four parameters,  $a_{B}$,  $b_{B}$,  $a_{B}^{\prime}$ and $b_{B}^{\prime}$, which are tabulated in Table~\ref{tab:QMC-parameter}.  
%%%%%%%%%%%%%%%%%%%%%%%%%%%%%%%%%%%%%%%%%%%%%%%%%%%%%%%%%%%%%%%%%%%%%%%%%%%%%%%
\begin{table}[t]
\caption{\label{tab:QMC-parameter}
Parameters %, $a_{B}$, $b_{B}$, $a_{B}^{\prime}$ and $b_{B}^{\prime}$, 
for the octet-baryon masses in the QMC and CQMC models.
}
\begin{tabular}{lcccccccc}
\hline
\         & \multicolumn{4}{c}{QMC}                                           & \multicolumn{4}{c}{CQMC}                                          \\
$B$       & $a_{B}$~(fm) & $b_{B}$ & $a_{B}^{\prime}$~(fm) & $b_{B}^{\prime}$ & $a_{B}$~(fm) & $b_{B}$ & $a_{B}^{\prime}$~(fm) & $b_{B}^{\prime}$ \\
\hline
$N$       & 0.179        & 1.00    & ---                   & ---              & 0.118        & 1.04    & ---                   & ---              \\
$\Lambda$ & 0.172        & 1.00    & 0.220                 & 1.00             & 0.122        & 1.09    & 0.290                 & 1.00             \\
$\Sigma$  & 0.177        & 1.00    & 0.223                 & 1.00             & 0.184        & 1.02    & 0.277                 & 1.15             \\
$\Xi$     & 0.166        & 1.00    & 0.215                 & 1.00             & 0.181        & 1.15    & 0.292                 & 1.04             \\
\hline
\end{tabular}
\end{table}
%%%%%%%%%%%%%%%%%%%%%%%%%%%%%%%%%%%%%%%%%%%%%%%%%%%%%%%%%%%%%%%%%%%%%%%%%%%%%%%
We note that the QMC and CQMC models can mostly explain the properties of nuclear matter around the nuclear saturation density without the NL potential.

To consider the relationship among the meson-baryon couplings,  it is extremely useful to use the SU(3)-invariant interaction Lagrangian.  
In this study,  we focus on the isoscalar,  vector-meson ($\omega$ and $\phi$) couplings to the octet baryons.  
Thus,  the coupling constants in SU(3) symmetry are expressed by 
\begin{eqnarray}
	g_{\omega\Lambda} = g_{\omega\Sigma}
	&=& \frac{1}{1+\sqrt{3}z\tan\theta_{v}} g_{\omega N} \ ,
	\hspace{0.5cm}
	g_{\omega\Xi}
	= \frac{1-\sqrt{3}z\tan\theta_{v}}{1+\sqrt{3}z\tan\theta_{v}} g_{\omega N} \ ,
	\label{eq:ESC-relations1} \\
	g_{\phi N}
	&=& \frac{\sqrt{3}z-\tan\theta_{v}}{1+\sqrt{3}z\tan\theta_{v}} g_{\omega N} \ ,
	\label{eq:ESC-relations2} \\
	g_{\phi\Lambda} = g_{\phi\Sigma}
	&=& \frac{-\tan\theta_{v}}{1+\sqrt{3}z\tan\theta_{v}} g_{\omega N} \ ,
	\hspace{0.5cm}
	g_{\phi\Xi}
	= -\frac{\sqrt{3}z+\tan\theta_{v}}{1+\sqrt{3}z\tan\theta_{v}} g_{\omega N} \ ,
	\label{eq:ESC-relations3}
\end{eqnarray}
with $z$ the coupling ratio,  $g_{8}/g_{1}$,  and $\theta_{v}$ the mixing angle of the vector meson.  
We here adopt $\theta_{v} = 37.50^{\circ}$ and $z = 0.1949$, which are given by the Nijmegen extended-soft-core model~\cite{Rijken:2010zzb}.  
If the {\it ideal} mixing angle,  $\theta_{v}^{ideal}\simeq 35.26^{\circ}$, and the value,  $z=1/\sqrt{6}\simeq 0.4082$,  are used,  
we can get the usual SU(6) relations based on the quark model.  

\section{Numerical Results}

\begin{figure}[t]
\includegraphics[width=188pt,keepaspectratio,clip,angle=270]{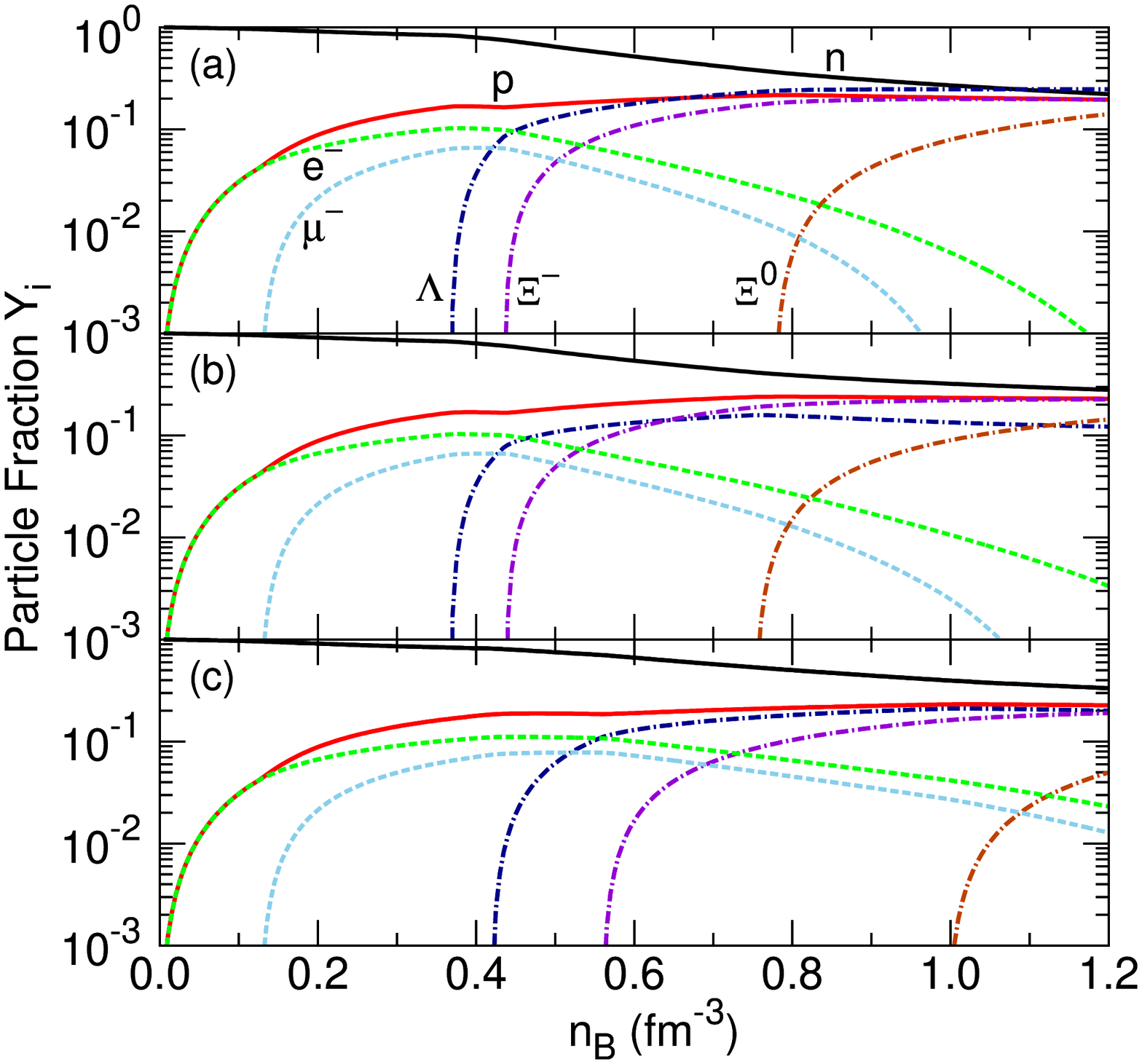}%
\hspace{0.5cm}
\includegraphics[width=188pt,keepaspectratio,clip,angle=270]{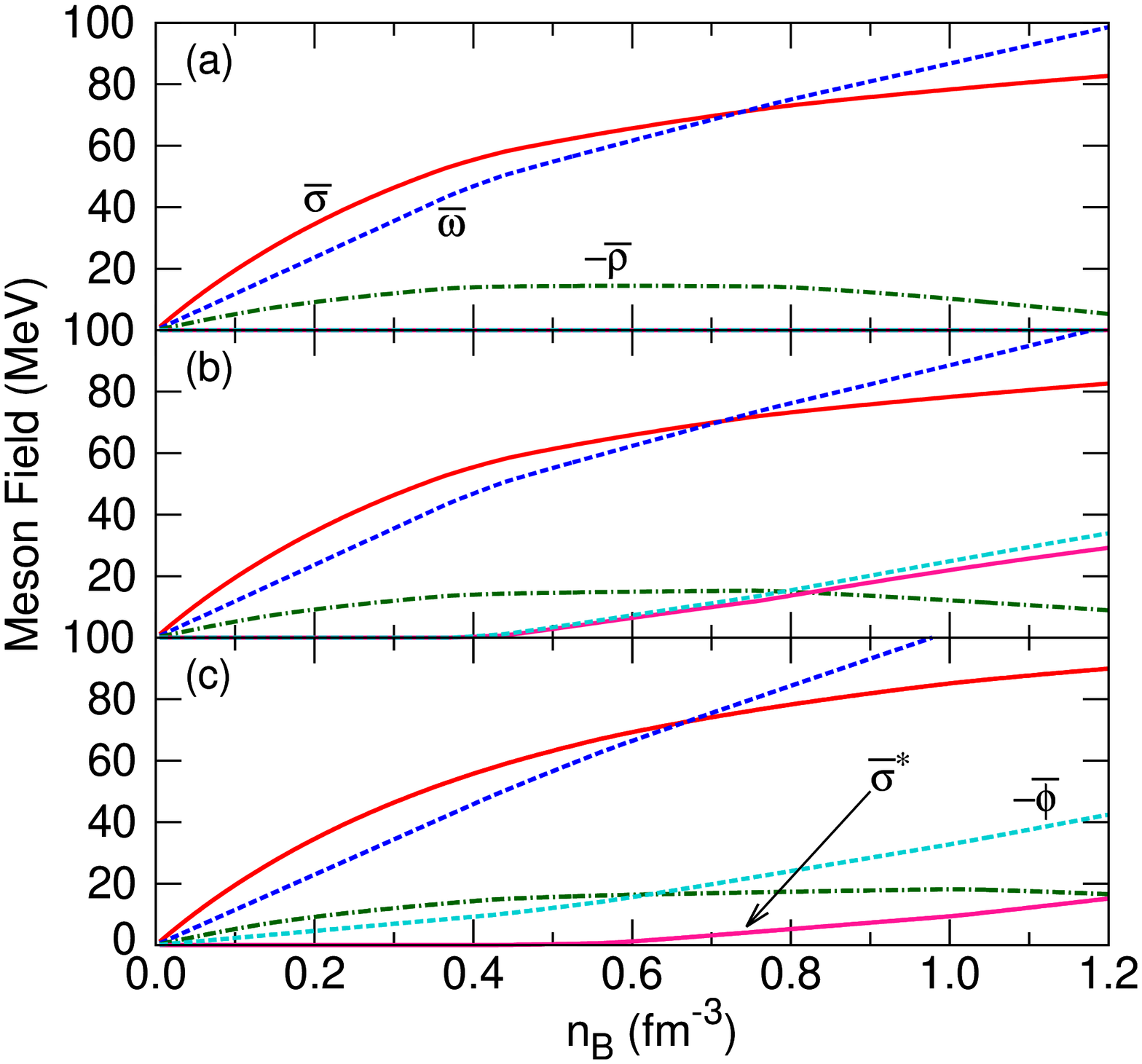}%
\caption{\label{fig:CQMC} 
Particle fractions,  $Y_{i}$,  (left panel) and meson fields (right panel) in the CQMC model.  
The constant mean-field values for the mesons are denoted by ${\bar \sigma}$, ${\bar \omega}$, etc.  
}
\end{figure}
\begin{figure}[t]
\includegraphics[width=143.5pt,keepaspectratio,clip,angle=270]{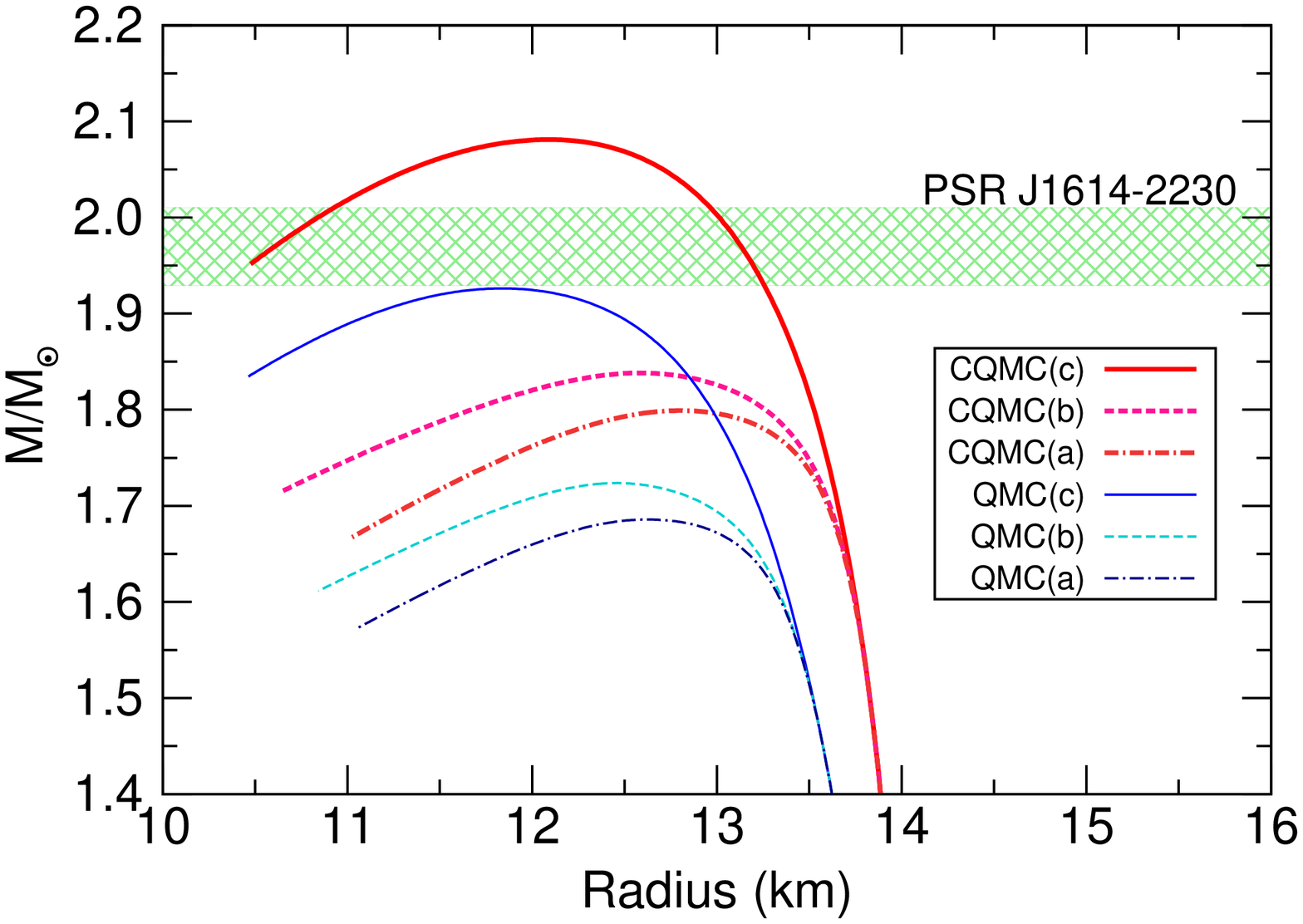}%
\hspace{0.5cm}
\includegraphics[width=143.5pt,keepaspectratio,clip,angle=270]{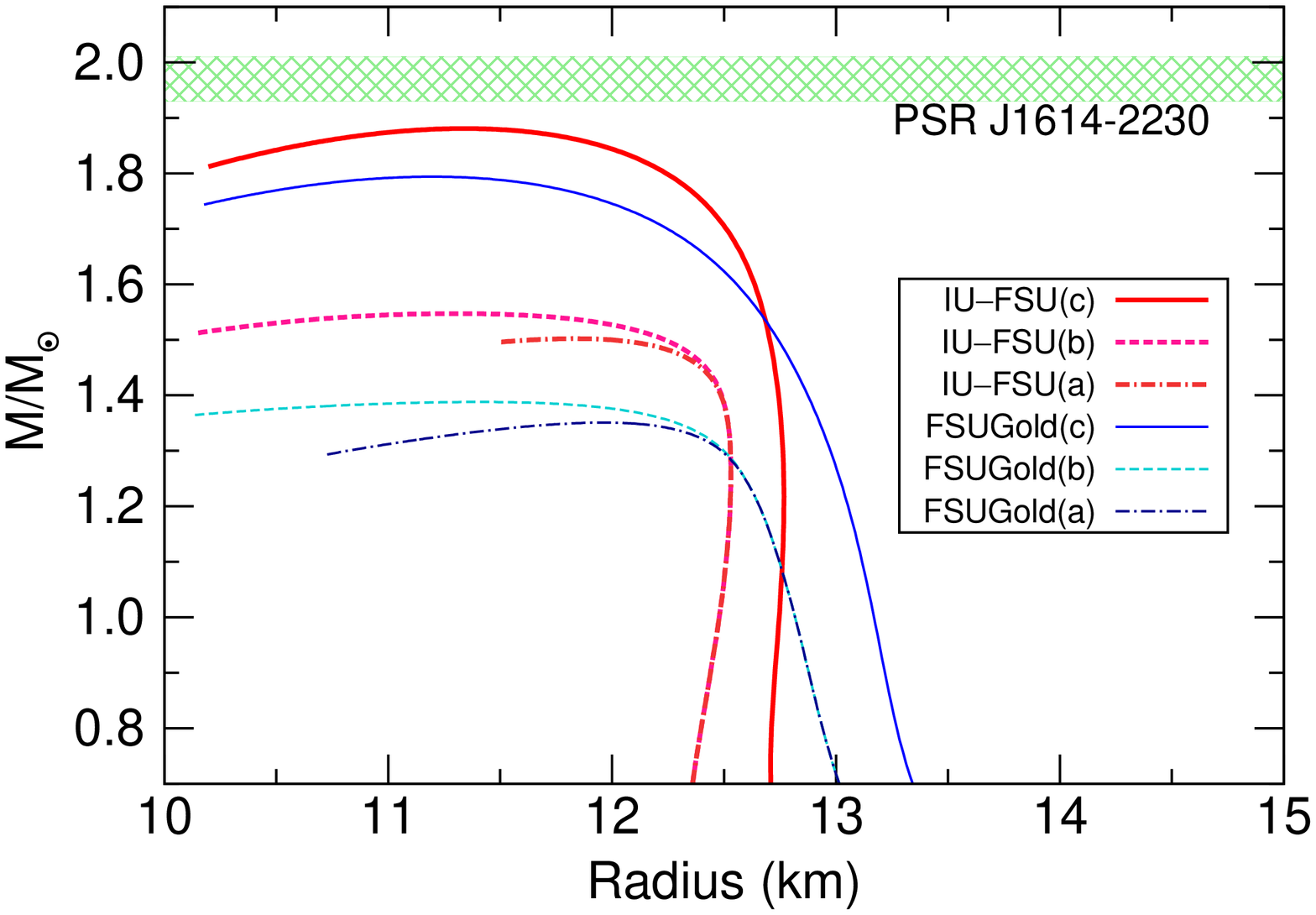}%
\caption{\label{fig:TOV} 
Mass-radius relations in the QMC and CQMC models (left panel) and in the FSUGold and IU-FSU models (right panel).  
The shaded area shows the mass of PSR J1614-2230,  $1.97 \pm 0.04 M_{\odot}$~\cite{Demorest:2010bx}.
}
\end{figure}

We adopt the parameter sets of the GM1, GM3, NL3, TM1, FSUGold and IU-FSU models as well as of the QMC and CQMC models.  
For more details, see Ref.~\cite{Miyatsu:2013yta} and references therein. 
In SU(3) flavor symmetry,  we readjust the coupling constants so as to reproduce the saturation condition in the original model.  
To study the properties of a neutron star,  we then solve the Tolman-Oppenheimer-Volkoff equation by using the EOSs calculated both in SU(6) and in SU(3) symmetries. 

In Fig.~\ref{fig:CQMC},  we present the particle fractions,  $Y_{i}$,  and meson fields in the CQMC model.  
In each panel,  the figure (a) is for the case where only the non-strange mesons ($\sigma$, $\omega$ and $\rho$) are considered in SU(6) symmetry,  the figure (b) is for the case where all the mesons including the $\sigma^{\ast}$ and $\phi$ are considered in SU(6) symmetry,  and the figure (c) is for the case where all the mesons are included in SU(3) symmetry.  
From (a) to (c) in order,  the threshold density for hyperon ($\Lambda$ or $\Xi^{0-}$) creation becomes higher.  
In SU(3) symmetry,  the $\phi$ meson contributes to the baryon interactions even at low densities because of the mixing effect.  

The mass-radius relations of neutron stars in the QMC and CQMC models and in the FSUGold and IU-FSU models are shown in Fig.~\ref{fig:TOV}.  
The inclusion of the strange mesons makes the EOS stiff,  and the maximum neutron-star mass is thus pushed upwards (from (a) to (b)).  
Furthermore,  the extension from SU(6) to SU(3) symmetry plays an important role in supporting a heavy neutron star (compare (b) with (c)).  

The properties of a neutron star in SU(6) or in SU(3) symmetry are listed in Table~\ref{tab:NS-properties}.  
The EOSs calculated by the QMC,  CQMC,  GM1 and TM1 in SU(3) symmetry can explain the mass of PSR J1614-2230,  $1.97 \pm 0.04 M_{\odot}$~\cite{Demorest:2010bx}.  
In the NL3 model with SU(3) symmetry,  the nucleon mass becomes negative before the neutron-star mass reaches the maximum point.  
Therefore,  the maximum mass in SU(3) symmetry is not given.  

%%%%%%%%%%%%%%%%%%%%%%%%%%%%%%%%%%%%%%%%%%%%%%%%%%%%%%%%%%%%%%%%%%%%%%%%%%%%%%%
\begin{table}[t]
\caption{\label{tab:NS-properties}
Properties of a neutron star in SU(6) or in SU(3) symmetry.  All the mesons ($\sigma$,  $\omega$,  $\rho$,  $\sigma^{\ast}$ and $\phi$) are considered.  
The neutron-star radius,  $R_{\max}$ (in km),  the ratio of the neutron-star mass to the solar mass,  $M_{\max}/M_{\odot}$,  and the central density,  $n_{c}$ (in fm$^{-3}$) at the maximum-mass point are listed.
}
\begin{tabular}{lcccccc}
\hline
\       & \multicolumn{3}{c}{SU(6)}                   & \multicolumn{3}{c}{SU(3)}                          \\
\       & $R_{\max}$ & $M_{\max}/M_{\odot}$ & $n_{c}$ & $R_{\max}$ & $M_{\max}/M_{\odot}$ & $n_{c}$        \\
\hline
QMC     & 12.5       & 1.72                 & 0.85    & 11.8       & 1.93                 & 0.96           \\
CQMC    & 12.6       & 1.84                 & 0.84    & 12.1       & 2.08                 & 0.90           \\
GM1     & 12.7       & 1.86                 & 0.82    & 12.2       & 2.14                 & 0.87           \\
GM3     & 12.1       & 1.63                 & 0.93    & 11.4       & 1.85                 & 1.05           \\
NL3      & 13.1       & 2.07                 & 0.78    & ---        & ---                  & ---            \\
TM1     & 13.1       & 1.72                 & 0.77    & 12.5       & 2.03                 & 0.86           \\
FSUGold & 11.4       & 1.39                 & 1.03    & 11.2       & 1.79                 & 1.08           \\
IU-FSU  & 11.3       & 1.55                 & 1.03    & 11.3       & 1.88                 & 1.02           \\
\hline
\end{tabular}
\end{table}
%%%%%%%%%%%%%%%%%%%%%%%%%%%%%%%%%%%%%%%%%%%%%%%%%%%%%%%%%%%%%%%%%%%%%%%%%%%%%%%

\section{Summary}

We have calculated the EOS for neutron stars,  using the popular RMF models as well as the QMC and CQMC models.  
As a result,  We have found that the models except GM3, FSUGold and IU-FSU can explain the masses of J1614-2230 and/or J0348+0432 in SU(3) symmetry.  
The extension from SU(6) to SU(3) symmetry and the strange vector meson,  $\phi$,  are very significant in sustaining massive neutron stars.  
In addition,  the variation of baryon structure in matter helps prevent the collapse of a neutron star.  

\section*{Acknowledgments}

This work was supported by the National Research Foundation of Korea (Grants No. 2012M7A1A2055605 and No. 2011-0015467).

\end{document}